\documentclass[aps,twocolumn,superscriptaddress,reprint,pre,showkeys]{revtex4-1}
\usepackage{bm}
\usepackage{graphicx}
\usepackage{amssymb}
\usepackage{amsmath}
\usepackage{physics}
\usepackage{float}
\usepackage{pgfplots}
\usepackage{placeins}
\usepackage{pifont}

\newcommand{\x}{\bm{x}}
\newcommand{\q}{\bm{q}} 
\newcommand{\Drho}{\Delta\rho}
\newcommand{\drho}{\delta\rho}
\newcommand{\hrho}{\hat{\rho}}

\begin{document}

\title{Improved field theoretical approach to noninteracting Brownian particles in a quenched random potential}

\author{Wonsang Lee}
\affiliation{Department of Physics, Konkuk University, Seoul 05029, Korea}
\author{Joonhyun Yeo}
\email{jhyeo@konkuk.ac.kr}
\affiliation{Department of Physics, Konkuk University, Seoul 05029, Korea}
\affiliation{School of Physics, Korea Institute for Advanced Study, Seoul 02455, Korea}

\date{\today}

\begin{abstract}
We construct a dynamical field theory for noninteracting Brownian particles in the presence of 
a quenched Gaussian random potential. 
The main variable for the field theory is the density fluctuation which measures the difference between the local density and its
average value. The average density is spatially inhomogeneous for given realization of the random potential. 
It becomes uniform only after averaged over the disorder
configurations. We develop the diagrammatic perturbation theory for the density correlation 
function and calculate the zero-frequency component of the response function exactly by summing all the diagrams contributing to it. 
From this exact result and the fluctuation dissipation relation, which holds in an equilibrium dynamics, 
we find that the connected density correlation function always decays to zero in the long-time limit for all values of disorder strength
implying that the system always remains ergodic.
This nonperturbative calculation relies on the simple diagrammatic structure of the present field theoretical scheme. 
We compare in detail our diagrammatic perturbation theory 
with the one used in a recent paper [B.\ Kim, M.\ Fuchs and V.\ Krakoviack, J.\ Stat.\ Mech.\ (2020) 023301],
which uses the density fluctuation around the uniform average, and discuss the difference in the
diagrammatic structures of the two formulations. 
\end{abstract}

%\pacs{, , }

% \noindent{\it Keywords\/}: Stochastic particle dynamics (Theory), Fluctuations (Theory), Stochastic processes (Theory)

\maketitle

\section{Introduction}

Dynamics of fluids in a quenched random environment has been studied
in connection with many different research areas ranging from the structural glass transition 
\cite{kim2003effects,cammarota2012ideal,karmakar2013random,krakoviack2011mode,konincks2017dynamics}
to biological \cite{hofling2013anomalous,meroz2015toolbox} and engineering 
\cite{king1987use,hristopulos2003renormalization,chen2000disorder,de1995correlation} applications. Theoretically the main focus
has been on the possible existence of an anomalous diffusion 
\cite{fisher1984random,visscher1984universality,kravtsov1985random,zhang1986diffusion,bouchaud1987anomalous,zwanzig1988diffusion,
tao1988exact,honkonen1988random,de1989invariance,deem1994classical,simon2013transport}, which has been studied 
in connection with the spatial dimension and the range of the random potential and the thermal noise
\cite{havlin1987diffusion,haus1987diffusion,bouchaud1990anomalous}.
One of the main physical quantities is the late time diffusion constant of a tagged particle. 
The calculation of effective transport properties in the presence of disorder has also been extensively studied
\cite{dean1994perturbation,dean1995perturbation,dean1996renormalization,romero1998brownian,
dean2004perturbation,dean2007effective,su2017colloidal}.

All these studies are, however, based on the single particle picture. 
An alternative way is to use the field theoretical approaches
\cite{das1985hydrodynamic,das1986fluctuating,schmitz1993absence,kawasaki1997path,miyazaki2005mode,
andreanov2006dynamical,kim2006mode,basu2007perspectives,kim2008fluctuation,jacquin2011field,kim2014equilibrium,velenich2008brownian},
which are based on the Martin-Siggia-Rose-Janssen-de Dominicis (MSRJD)-type dynamical field theory  \cite{martin1973statistical,janssen1976lagrangean,de1978dynamics} for the stochastic equations governing collective variables
such as the density field.
There are a number of advantages of using the field theoretical formalism. One can for example 
extract physical information from the symmetry property of the action functional. For instance, the fluctuation dissipation relation (FDR)
of an equilibrium dynamics is obtained from the invariance of the action functional 
under time-reversal transformations of fields \cite{andreanov2006dynamical,aron2010symmetries}. 
Another point is the availability of systematic techniques such as the loop expansion and 
the diagrammatic resummation methods in the field theoretical setting.
In fact, there have been a series of attempts \cite{andreanov2006dynamical,kim2006mode,
basu2007perspectives,kim2008fluctuation,jacquin2011field,kim2014equilibrium} to obtain the 
mode coupling theory (MCT) \cite{gotze2008complex} of the structural glass transition 
as a first-order self-consistent renormalized theory from the field theoretical formulation of dense liquids or colloids. 

There are nontrivial technical issues arising in the field theoretical formalism.
For colloidal fluids described by Brownian particles obeying overdamped Langevin equations for the particle positions,
an alternative stochastic equation for the microscopic density known as the Dean-Kawasaki (DK)
equation \cite{kawasaki1994stochastic,dean1996langevin} can be obtained. The field theoretical formulation of 
the DK equation even in the simplest case of noninteracting Brownian particles turns out to be nontrivial \cite{velenich2008brownian}.
Even in the absence of interaction among particles, the field theory contains an interaction originated from the multiplicative noise in the DK equation.
The multiplicative noise is also responsible for the nontrivial form of the response function \cite{miyazaki2005mode,andreanov2006dynamical}
which appears in the FDR. Recently, the effect of quenched disorder on the field theoretical formulation of 
noninteracting Brownian particles was studied \cite{kim2020dynamics}. The quenched disorder produces another form of interaction
in addition to the one due to the multiplicative noise. Various schemes of perturbation and self-consistent renormalization 
theories consistent with the FDR were developed. It was shown that for some renormalized perturbation scheme the system
becomes nonergodic for strong disorder in the sense that the connected correlation function 
does not decay to zero in the long time limit, while still exhibiting the normal diffusion \cite{kim2020dynamics}. 

In this paper, we present an improved field theoretical formulation compared to that in Ref.~\cite{kim2020dynamics}
for the noninteracting Brownian particles
in a quenched random potential with a Gaussian correlation. 
The main variable in the field theory is the density fluctuation, which measures the difference of 
the local density from its average value, which is spatially inhomogeneous depending on a
given realization of the random potential. Only after averaged over the disorder, a uniform average density emerges.
We present the field theoretical formulation in terms of the density fluctuation around its inhomogeneous average value.
In Ref.~\cite{kim2020dynamics}, the field theory for the same problem was formulated in terms of 
the density fluctuation with respect to the uniform average value.  We find that although the two versions of the field theory
differ only by the definition of the main variable and eventually
give the same physical quantities, the detailed diagrammatic perturbation 
theories are quite different. 
We show that within the formalism of Ref.~\cite{kim2020dynamics}, 
the perturbation expansion in terms of the disorder strength 
produces a multitude of unnecessary diagrams which either cancel among themselves or vanish.
In our version of the field theory, however, the diagrammatic expansion has a much simpler structure.
As an example of the simplicity, we show that we can evaluate the zero-frequency limit or the time integral of 
the response function exactly by summing all the diagrams contributing to it.
This quantity gives, via the FDR, the value of the density-density correlation function in the long time limit. 
From the exact calculation, we find that the system remains ergodic for all values of the disorder strength
in the sense that the connected density-density correlation function always decays to zero in the long time limit.

This paper is organized as follows. In Sec.~\ref{sec:field}, 
we present the MSRJD field theory for the noninteracting Brownian particles in the presence of 
the quenched random potential. In the next section, we compare the diagrammatic perturbation theory
of our formulation with that used in Ref.~\cite{kim2020dynamics}. 
In Sec.~\ref{sec:resp}, we calculate the zero frequency limit of the response function and
discuss its implication on the ergodicity of the system. We conclude in the following section with discussion.

%\vspace{3mm}
\section{MSRJD Field Theory for noninteracting Brownian particles in a quenched random potential}
\label{sec:field}
%\subsection{Model}
%\subsection{MSRJD formalism for the density fluctuation}
We consider $N$ noninteracting Brownian particles moving in $d$ dimensions under the influence of 
an external random potential $\Phi$. 
The position $\bm{X}_i(t)$ of the $i$-th particle at time $t$
is described by the overdamped Langevin equations,
\begin{equation}
\frac{d\bm{X}_{i}(t)}{dt}=-\Gamma\bm{\nabla}\Phi(\bm{X}_{i}(t)) +\bm{\eta}_{i}(t), \label{originalmodel}
\end{equation}
where  $\bm{\eta}_{i}(t)$ 
is the Gaussian white noise having zero mean and variance
\begin{equation}
 \langle \eta^\mu_i(t)\eta^\nu_j(t^\prime)\rangle =2D\delta_{ij}\delta^{\mu\nu}\delta(t-t^\prime)
\end{equation}
for $i,j=1,\ldots,N$ and $\mu,\nu=1,\ldots,d$ and $D=\Gamma T$ with the temperature $T$ ($k_B=1$). 
The quenched random potential $\Phi(\bm{x})$ is selected from the Gaussian 
distribution with zero mean and the variance
\begin{equation}
\overline{\Phi(\x_1)\Phi(\x_2)}=\Delta(\x_1-\x_2), \label{phiphi}
\end{equation}
where $\Delta$ is a short-ranged function of the distance between the two points. Here
the overline indicates the average over the quenched random potential. 

In this paper, we study this system using the microscopic density field
\begin{equation}
\rho(\x,t)= \sum _{i=1}^{N}\delta^{(d)}(\x-\bm{X}_{i}(t)),
\end{equation}
as a main dynamical variable.
The above equation can then be rewritten as a stochastic equation with a 
multiplicative noise \cite{kawasaki1994stochastic,dean1996langevin} as
\begin{align}
\label{DeanEquation}
\frac{\partial\rho(\x,t)}{\partial t}=\Gamma & \nabla \cdot \left \{ \rho(\x,t) \nabla\Phi(\x) \right \} +  D\nabla^{2}\rho(\x,t) \nonumber \\
&- \nabla\cdot \left \{ \bm{\xi}(\x,t)   \sqrt{\rho(\x,t)} \right \} ,
\end{align}
where 
\begin{equation}
\left \langle \xi^{\mu}(\x,t)\xi^{\nu}(\x',t')  \right \rangle = 2D \delta^{\mu\nu} \delta(t-t')\delta^{(d)}(\x-\x').
\end{equation}
We note that in the above derivation, Ito's discretization convention was used. 

Many physical quantities such as the intermediate scattering functions are obtained from the correlation
functions of density fluctuation. For given realization of the external random potential, 
we consider the fluctuation $\delta\rho(\bm{x})$ around its average value
$\langle \rho(\x,t) \rangle$, which we denote by $\rho_{\Phi}(\x)$:
\begin{equation}
\label{DrhoDef}
\drho(\x,t) \equiv  \rho(\x,t) - \rho_{\Phi}(\x).
\end{equation}
In the presence of the external potential, $\rho_\Phi(\bm{x})$ will be inhomogeneous and can be determined 
from the stationarity condition of Eq.~(\ref{DeanEquation}) as
\begin{equation}
\nabla \cdot \left \{ \rho_{\Phi}(\x) \nabla\Phi(\x) \right \} +  D\nabla^{2}\rho_{\Phi}(\x)=0
\end{equation}
with the solution 
\begin{equation}
\rho_{\Phi}(\x)=N \frac{e^{-\Phi(\x)/T}}{\int d^d\x\; e^{-\Phi(\x)/T}}. \label{rhophi}
\end{equation}
We can show that the mean and the variance of the denominator on the right hand side of Eq.~(\ref{rhophi})
over the Gaussian distribution of the random potential
are both proportional to the volume (see Appendix \ref{app:a}). 
Therefore, in the thermodynamic limit, we can replace it by its disorder average value,
$\overline{\int d^d\x e^{-\Phi(\x)/T}} =V e^{\Delta(\bm{0})/(2T^2)}$ with the volume V. 
This is consistent with the fact that when averaged over the disorder realizations, we have 
$\overline{\rho_\Phi(\bm{x})}=  \rho_0 \equiv N / V  $.  
For future use, we rewrite Eq.~(\ref{rhophi}) as
\begin{equation}
\rho_{\Phi}(\x)=\rho^\ast_0 e^{-\Phi(\x)/T} , \label{rhophi_2}
\end{equation}
where $\rho^\ast_0\equiv \rho_0 e^{-\Delta(\bm{0})/(2T^2)}$.
We shall use $\drho $ defined in Eq.~(\ref{DrhoDef}) as the main variable for the MSRJD field theoretical development below. 
An alternative way is to use the density fluctuation around its uniform average as 
\begin{equation}
\label{DifferentDrhoDef}
\Drho(\x,t) \equiv \rho(\x,t)-\rho_0.
\end{equation}
The field theoretical formulation using $\Delta\rho$ will be discussed in the next section in detail.

We write $\rho=\rho_\Phi+\delta\rho$ in Eq.~(\ref{DeanEquation}) and transform it 
into a field theoretical setting using the MSRJD formalism \cite{martin1973statistical,janssen1976lagrangean,de1978dynamics},
for which the generating functional is written as path integrals over the density fluctuation $ \drho(\x,t)$ and the auxiliary response field $\hrho(\x,t)$.
For given external potential $\Phi$, the average of an observable $O(\drho,\hrho)$ is then given by the functional integral
\begin{equation}
\langle O(\drho,\hrho) \rangle =\int  \mathcal{D}\drho \int \mathcal{D}\hrho  \;   O(\drho,\hrho) \; e^{S_\Phi[\drho,\hrho]},
\end{equation} 
where
\begin{widetext}
\begin{align}
S_\Phi[\drho,\hrho]=\int d^d\x \int dt &\biggl [- i\hrho(\x,t)\left[ \frac{\partial\drho(\x,t)}{\partial t}-D\nabla^{2}\drho(\x,t)  
-\Gamma \nabla\cdot\left \{ \drho(\x,t)\nabla\Phi(\x) \right \}  \right] \nonumber \\
&+D\rho_{\Phi}(\x)\left \{ \nabla i\hrho (\x,t) \right \}^{2} 
+ D\drho(\x,t)\left \{\nabla i\hrho(\x,t) \right \}^{2} \biggr ] . \label{SPhi}
\end{align}
The average over the disorder realization can be obtained by integrating it over the distribution 
\begin{equation}
\mathcal{P}[\Phi]=\frac 1 {\mathcal{N}} \exp[-  \frac{1}{2}\int d^d\x \int d^d\x' \Phi(\x)\Delta^{-1}(\x-\x')\Phi(\x')]
\end{equation}
for the external potential, where $\mathcal{N}$ is the normalization constant and 
$ \Delta^{-1} $ is the matrix inverse of $ \Delta $. Therefore we have
\begin{equation}
\overline{\langle O(\drho,\hrho) \rangle} 
 =\int  \mathcal{D}\drho \int \mathcal{D}\hrho \int [\mathcal{D}\Phi ] \; 
 O(\drho,\hrho) \; e^{S[\drho,\hrho,\Phi]} ,
\end{equation}
where the functional integral $[\mathcal{D}\Phi]$ contains the normalization factor
$1/\mathcal{N}$ and the effective action $S$ is given by
\begin{align}
\label{S_Phi_Field_d}
S[\drho,\hrho&,\Phi] \equiv S_\Phi[\drho,\hrho] -  \frac{1}{2}\int d^d\x \int d^d\x' \Phi(\x)\Delta^{-1}(\x-\x')\Phi(\x') .
\end{align}

If we expand Eq.~(\ref{rhophi_2}) in powers of $\Phi$, 
we can separate the effective action into the Gaussian and non-Gaussian part as $S=S_{G}+S_{v}$ where
\begin{equation}
\label{EffectiveActionG}
S_G=\int d^d\x \int dt  \left [- i\hrho(\x,t)\left\{  \frac{\partial \drho(\x,t)}{\partial t}-D \nabla ^{2}\drho( \x,t)\right\} 
+D\rho^\ast_0 (\nabla i\hrho(\x,t) )^{2} \right] -  \frac{1}{2}\int d^d\x\int d^d\x' \Phi(\x)\Delta^{-1}(\x-\x')\Phi(\x') ,
\end{equation}
and
\begin{equation}
\label{EffectiveActionnG}
\begin{split}
S_{v}=\int d^d\x \int dt\left [D\drho(\x,t) ( \nabla i\hrho(\x,t) )^{2} 
+\Gamma ( i\hrho(\x,t)) \nabla\cdot\left \{ \drho(\x,t)\nabla\Phi(\x) \right \} + D \rho^\ast_0 (\nabla i\hrho(\x,t) )^{2}
\sum_{n=1}^{\infty} \frac{\Phi^n(\x)}{n!(-T)^n}
   \right] .
\end{split}
\end{equation}
\end{widetext}
In the absence of the random potential ($\Phi=0$ and $\Delta=0$), we recover the action studied in 
Ref.~\cite{velenich2008brownian} for the Brownian gas. 
The first term in Eq.~(\ref{EffectiveActionnG}), which we refer to as the noise vertex, comes from the multiplicative noise
and its effect has been studied in
the field theoretical setting in the absence of disorder in Ref.~\cite{velenich2008brownian}. 
The second and third
terms in Eq.~(\ref{EffectiveActionnG}) are the vertices arising from the quenched random potential. 
The second term is linear in $ \Phi $, but the third term is an infinite series of terms which originate from
the expansion of $\rho_\Phi$.
We denote these as  the type-$\mathcal{A}$ and type-$\mathcal{B}_n$ vertices ($n=1,2,\cdots$), respectively.
Below we study the effects of these vertices on the perturbation theory in detail.

The central quantities in this paper are the two-point functions defined by
\begin{equation}
G_{\alpha\beta}(\x-\x',t-t')= \langle \psi_\alpha(\x,t)\psi_\beta(\x',t')\rangle_S,
\label{2pt}
\end{equation}
where $\psi_\alpha$ and $\psi_\beta$ represent $\hat{\rho}$ or $\drho$
and the average $\langle\cdots\rangle_S$ is with respect to the effective action $S$
in Eqs.~(\ref{EffectiveActionG}) and (\ref{EffectiveActionnG}). 
(In the subscript, we use $\rho$ instead of $\delta\rho$ for brevity.)
Their Fourier transforms are given by
\begin{equation}
 \tilde{G}_{\alpha\beta}(\q,t)=\int d^d\x \; e^{-i\q\cdot\x}G_{\alpha\beta}(\x,t).
\end{equation}
We first note that due to causality $G_{\hat{\rho}\hat{\rho}}=0$.

\begin{figure}
  \includegraphics[width=0.32\columnwidth]{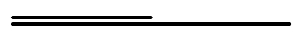} 
  \includegraphics[width=0.32\columnwidth]{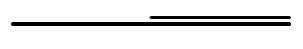} 
  \includegraphics[width=0.32\columnwidth]{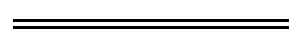} 
  \caption{Diagrams for the bare propagators, $i\tilde{G}^0_{\rho\hat{\rho}}(\q,t)$, $i\tilde{G}^0_{\hat{\rho}\rho}(\q,t)$
  and $\tilde{G}^0_{\rho\rho}(\q,t)$ from left to right. Double lines are for $\delta\rho$ and single lines for $i\hat\rho$.  }
  \label{bare_prop}
\end{figure}

The bare propagators are obtained from the Gaussian action in Eq.~(\ref{EffectiveActionG}). These will be denoted by 
$G^0_{\psi_\alpha\psi_\beta}$ and by the lines shown in Fig.~\ref{bare_prop}.
The double lines indicate $\delta\rho$ and the single lines are $i\hat\rho$. The
bare propagators are given by
\begin{align}
\label{bare_prop_l}
&i\tilde{G}^0_{\rho\hat{\rho}}(\q,t)= \Theta(t) e^{-D\q^{2}t}, \\
\label{bare_prop_r}
&i\tilde{G}^0_{\hat{\rho}\rho}(\q,t) = \Theta(-t) e^{D\q^{2}t}, \\
\label{bare_prop_c}
&\tilde{G}^0_{\rho\rho}(\q,t)=\rho^\ast_0 e^{ -Dq^{2}|t |},
\end{align}
where $\Theta(t)$ is the step function. 
The bare propagator involving $ \Phi $ is $ G^0_{\Phi\Phi}(\x-\x^\prime) = \Delta(\x-\x') $ of which Fourier transform is given by
\begin{equation}
 \tilde{\Delta}(\q)=\int d^d\x \; e^{-i\q\cdot\x}\Delta(\x).
\end{equation}
It will be denoted by dashed lines in diagrams, which carry momentum but not frequency.

\begin{figure}
  \includegraphics[width=0.30\columnwidth]{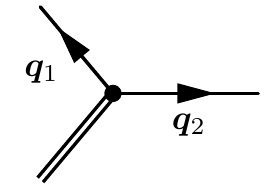} \;\;\;\;\;
  \includegraphics[width=0.30\columnwidth]{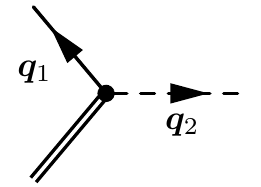} 
  \caption{Diagrams for the noise vertex (left), type-$\mathcal{A}$ vertex (right) }
  \label{vertices1}
\end{figure}

\begin{figure}
  \includegraphics[width=0.36\columnwidth]{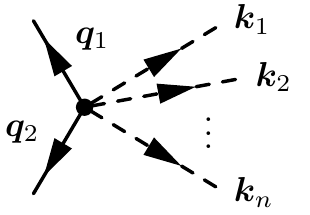} 
  \caption{Diagram for the type-$\mathcal{B}_n$ vertex.}
  \label{vertices2}
\end{figure}

The nonlinear vertices in the action, Eq.~(\ref{EffectiveActionnG}) are represented graphically as
in Figs.~\ref{vertices1} and \ref{vertices2}. At each black dot, the momentum conservation holds. 
The cubic noise vertex involves two $i\hat{\rho}$ fields and one $\delta\rho$ field
and is shown on the left panel of Fig.~\ref{vertices1}. It contributes the factor 
of $-D\bm{q}_{1}\cdot\bm{q}_{2}$ and does not 
involve the random potential. The type-$\mathcal{A} $ vertex is shown on the right panel of Fig.~\ref{vertices1}. It gives the factor of 
$\Gamma \q_1 \cdot \q_2$.
Finally, there are infinite number of vertices, labelled as type-$ \mathcal{B}_n $, 
as shown in Fig.~\ref{vertices2}, each of which carries the factor of
\begin{equation}
\label{typeB_n}
- \rho^\ast_0   \frac{D \q_1 \cdot \q_2}{n!(-T)^n}.
\end{equation}


\begin{figure}
  \includegraphics[width=0.40\columnwidth]{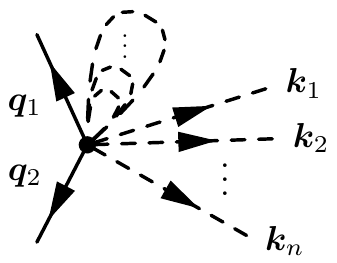} 
  \caption{Diagram representing the renormalized type-$\mathcal{B}_n$ vertex.}
  \label{vertices_renom}
\end{figure}

The correlation functions are obtained by connecting the lines in the vertices using the bare propagators.
Before going into this discussion, however, we first note that the $ \Delta $-dependent part in the average density 
$\rho^\ast_0$ appearing in Eqs.~(\ref{EffectiveActionG}), (\ref{EffectiveActionnG}), (\ref{bare_prop_c}) 
and (\ref{typeB_n}) can be eliminated
by considering the renormalized type-$\mathcal{B}_n$ vertices in the following way. 
We first consider the type-$\mathcal{B}_{n+2m}$ vertex for given $n,m=1,2,\cdots$. 
If we connect the $m$ pairs of dashed lines in the vertex as in Fig.~\ref{vertices_renom}, we end up with another 
type-$\mathcal{B}_n$ vertex. Since there are $(n+2m)!/(2^m n! m!)$ distinct ways of doing this, this new type-$\mathcal{B}_n$ vertex
carries a factor of
\begin{equation}
-  \rho^*_0  \frac{D \q_1 \cdot \q_2 \Delta^{m}(\bm{0})  }{2^m n! m!  (-T)^{n+2m}}.
\label{typeB_n_ex}
\end{equation}
Now if we sum all these contributions for $m=1,2,\cdots$ and combine with the bare vertex in Eq.~(\ref{typeB_n}),
which corresponds to $m=0$ case, we have
\begin{equation}
-  \rho^*_0 \frac{D \q_1 \cdot \q_2}{n! (-T)^n} \sum_{m=0}^\infty\frac{ \Delta^{m}(\bm{0})  }{m! 2^m   (-T)^{2m}} = - \rho_0   \frac{D \q_1 \cdot \q_2}{n!(-T)^n}.
\label{typeB_renom}
\end{equation}
We note that the effect of this renormalization is just the use of $\rho_0$ instead of $\rho^*_0$ in the type-$\mathcal{B}_n$ vertex.
This kind of discussion also applies to the third term in the Gaussian action in Eq.~(\ref{EffectiveActionG}), which can be regarded as the 
$n=0$ case. If we sum over all these self-pairing contributions from the type-$\mathcal{B}_{2m}$ vertices, we end up with the renormalized 
term which again carries $\rho_0$ instead of $\rho^\ast_0$. 
The renormalization of the Gaussian action therefore results in the renormalized bare propagator 
\begin{align}
\label{bare_prop_c_renom}
&\tilde{G}^0_{\rho\rho}(\q,t)=\rho_0 e^{ -Dq^{2}|t |}.
\end{align}
with just $\rho_0$.
Note that for this renormailzed bare propagator, we have
$\tilde{G}^0_{\rho\rho}(\q,t)=\rho_0[i\tilde{G}^0_{\rho\hat{\rho}}(\q,t)+i\tilde{G}^0_{\hat{\rho}\rho}(\q,t)]$, which we will use 
throughout this paper. In the perturbation theory developed below, we 
regard that the pairings of the dashed lines originating 
from a single vertex as done in Fig.~\ref{vertices_renom} are already taken care of by
using the renormalized type-$\mathcal{B}_n$ vertex. 

We now investigate the diagrams contributing to the two-point functions, Eq.~(\ref{2pt}) in the perturbation theory.
An interesting feature of the field theory constructed from these propagators and vertices is
that the number of the single lines ($\hat\rho$) in every vertex is greater than or equal to that of the double lines ($\delta\rho$). 
Since the correlation functions are obtained by pairing up these lines and 
since $G^0_{\hat{\rho}\hat{\rho}}=0$, every nonzero diagram should contain equal or more number of $\drho $ than $\hrho$. 
For $\tilde{G}_{\rho\hat{\rho}}(\q,t)$ and $\tilde{G}_{\hat{\rho}\rho}(\q,t)$, 
any combinations of the noise vertex with two $\hrho$'s and a $\drho$ and
the type-$ \mathcal{B}_n $ vertices with two $\hrho$'s do not contribute,
since they all have more $\hat{\rho}$'s than $\delta\rho$'s.  For this type of correlation function, therefore,
only the type-$\mathcal{A} $ vertex given in Fig.~\ref{vertices1} contributes. 

\begin{figure}
  \includegraphics[width=0.20\columnwidth]{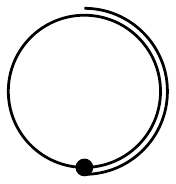} 
  \includegraphics[width=0.20\columnwidth]{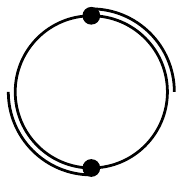} 
  \includegraphics[width=0.20\columnwidth]{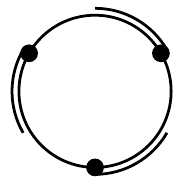} 
  \includegraphics[width=0.20\columnwidth]{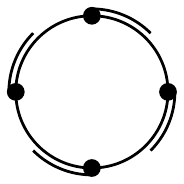} \;\;\;\;
\caption{Possible loops in the field theory. They all vanish due to causality.}
\label{loop}
\end{figure}

For $\tilde{G}_{\rho\rho}(\q,t)$, in addition to the type-$\mathcal{A} $ vertex, the type-$ \mathcal{B}_n $ vertices may contribute, since the two external $\delta\rho$'s can make up for the two extra $\hat{\rho}$'s in that vertex. 
In terms of the number of $\hat{\rho}$'s and $\drho$'s, the noise vertex may appear to  
contribute to $\tilde{G}_{\rho\rho}(\q,t)$. This is, however, not the case. 
Since there are only two external lines for $\tilde{G}_{\rho\rho}(\q,t)$ and since the noise
vertex has three legs, the only way 
for the noise vertex to contribute to $\tilde{G}_{\rho\rho}(\q,t)$ is to form a closed loop. 
But the closed loops that appear in this field theory are all in the form of response loops described in 
Fig.~\ref{loop}, since every vertex has only one double line ($\drho$) at most. These loop diagrams, however, vanish 
due to the causal structure of $\tilde{G}^0_{\rho\hat{\rho}}(\q,t)$ and $\tilde{G}^0_{\hat{\rho}\rho}(\q,t)$.
The absence of closed loops is an important feature of the present field theory, which has already been
noted in Ref.~\cite{velenich2008brownian} for the Brownian gas without disorder. 

\begin{figure}
\includegraphics{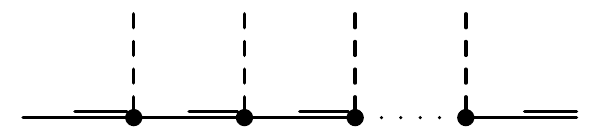} 
\caption{Construction of a generic diagram contributing to $ G_{\rho\hrho} $. 
There are even number of dots. 
Each diagram is generated by connecting all possible $n$ pairs of the dashed lines.}
\label{RespGen}
\end{figure}

For the functions $ G_{\rho\hrho} $ and $ G_{\hrho\rho} $, as discussed already, 
only the type-$\mathcal{A} $ vertex is relevant. 
Since there are no loops with respect to the single and double solid lines,
all the diagrams are tree diagrams with the connected dashed lines representing the disorder strength $\tilde{\Delta}(\bm{q})$.
We can present a general recipe for constructing a generic diagram contributing to, say 
$ G_{\hrho\rho} $ at an arbitrary order as in Fig.~\ref{RespGen}. 
We first put $ 2n $ dots between the external lines and make a single line by connecting 
them all with the bare $G^0_{\hat{\rho}\rho}$'s. 
We then connect $ n $ pairs of the dots using the dashed lines in all possible ways to generate a general diagram 
contributing to $ G_{\hrho\rho} $.

\begin{figure}
\includegraphics{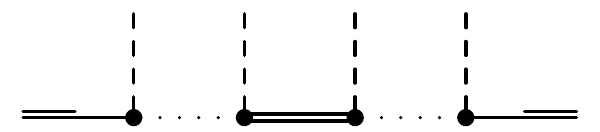} \\[8pt]
\includegraphics{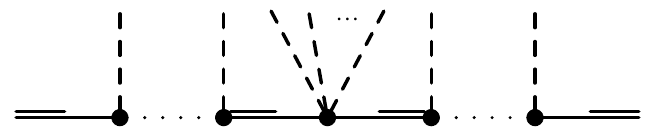} 
\caption{Two possible structures of diagrams contributing to $ G_{\rho\rho} $:
(a) diagrams without type-$ \mathcal{B}_n $ vertex (above) and
(b) diagrams containing one type-$ \mathcal{B}_n $ vertex (below).
As in Fig.~\ref{RespGen}, the dashed lines are paired up and connected
in all possible ways.}
\label{CorlGen}
\end{figure}

For $G_{\rho\rho}$,
by counting the number of $\rho$'s and $\hrho$'s, we note that
the type-$ \mathcal{B}_n $ vertices can only appear once at most in a diagram contributing to this function. 
we can classify the diagrams contributing to $ G_{\rho\rho} $ into two distinct categories: 
(a) those in which the type-$ \mathcal{B}_n $ vertices is not used, and (b) those where it is used only once. 
The structures of these two kinds of diagrams are described in Fig.~\ref{CorlGen}. 
They are constructed in a similar way to the previous case by connecting all possible pairs of dashed lines.

It is now a straightforward matter to develop a perturbation expansion in powers of $\Delta$. We have carried out 
the calculation of $G_{\rho\hat{\rho}}(\bm{q},t)$ and $G_{\rho\rho}(\bm{q},t)$ to the first order of $\Delta$. We find that 
in the presence of disorder, these functions decay much slower than the exponential decay exhibited by the bare functions.
In fact, we find that $G_{\rho\hat{\rho}}(\bm{q},t)$ and $G_{\rho\rho}(\bm{q},t)$ behave in the long time limit as
$t^{-d/2-1}$ and $t^{-d/2}$, respectively. The latter behavior was already observed in Ref.~\cite{kim2020dynamics}.
In this paper, rather than continuing the perturbation theory calculation to higher orders, we focus
on a nonperturbative calculation which will be presented in Sec.~\ref{sec:resp}. 

%\vspace{3mm}

\section{Field theory for the density fluctuation around the uniform average value}
\label{Sec_uav}

Before going to the nonperturbative calculation, we study in this section the field theory using the density fluctuation $\Delta\rho(\bm{x},t)$
around the uniform average density $\rho_0$, which is defined in Eq.~(\ref{DifferentDrhoDef}). 
This is the formalism used in Ref.~\cite{kim2020dynamics}. We present a
detailed comparison of the field theory with respect to this variable with the one developed 
in the previous section.
We will show that both formalisms give the same correlation functions and physical quantities as expected.
Nevertheless, we find it useful to compare these two formalisms
in detail for general understanding of the diagrmmatics involved in these field theories
and for the nonperturbative calculation which will be done in Sec.~\ref{sec:resp}.
In fact, we find that 
the perturbative field theory using $\Delta\rho$ 
is much more complicated than that for $\delta\rho$.

For given realization of the disorder potential,
the thermal average $\langle\Delta\rho(\bm{x},t)\rangle=\rho_{\Phi}(\bm{x})-\rho_0\equiv \delta\rho_\Phi(\bm{x})$ does not vanish. 
As we have seen in Eq.~(\ref{rhophi_2}),
only after averaged over the disorder potential, this quantity vanishes, $\overline{\langle\Delta\rho(\bm{x},t)\rangle}=0$. 
Higher order moments of $\Delta\rho$'s are related to those of $\delta\rho$'s, since we can write
$\Delta\rho(\bm{x},t)=\delta\rho(\bm{x},t)+\delta\rho_\Phi(\bm{x})$.
For example, we have \cite{konincks2017dynamics,kim2020dynamics}
\begin{widetext}
\begin{align}
\label{DeltaRho2Point}
\overline{\langle \Delta\rho(\bm{x},t)\Delta\rho(\bm{x}^\prime,t^\prime)\rangle}
=\overline{\langle \delta\rho(\bm{x},t)\delta\rho(\bm{x}^\prime,t^\prime)\rangle} 
+\overline{\delta\rho_\Phi(\bm{x})\delta\rho_\Phi(\bm{x}^\prime)},
\end{align}
where we can evaluate explicitly the last term as
\begin{equation}
\label{DisconnectedCorrelation}
\overline{\delta\rho_\Phi(\x)\delta\rho_\Phi(\x^\prime)} =\rho_0^2 \left(e^{\Delta(\x-\x^\prime)/T^2}-1 \right).
\end{equation}
The first term on the right hand side of Eq.~(\ref{DeltaRho2Point}) is just 
$G_{\rho\rho}(\bm{x}-\bm{x}^\prime, t-t^\prime)$ defined in the previous section.
Therefore,  
the correlation function on the left hand side of Eq.~(\ref{DeltaRho2Point}) differs from $G_{\rho\rho}(\bm{x}-\bm{x}^\prime, t-t^\prime)$ 
by the time-independent quantity given by Eq.~(\ref{DisconnectedCorrelation}).

We write $\rho=\rho_0+\Delta\rho$ in Eq.~(\ref{DeanEquation}) and again transform it into a field theory as done before.
The average of an observable $O(\Drho,\hrho)$ is then given by
\begin{equation}
\label{EffectiveActionAvg_naive}
\overline{\langle O(\Drho,\hrho) \rangle} =\int  \mathcal{D}\Drho \int\mathcal{D}\hrho \int[\mathcal{D}\Phi ]\;   O(\Drho,\hrho) \; e^{S_\Delta[\Drho,\hrho,\Phi]},
\end{equation}
where $S_\Delta=S^G_\Delta+S^v_\Delta$ with
\begin{equation}
\label{EffectiveAction_naive_G}
S^G_\Delta=\int d^d\x \int dt\Big[- i\hrho(\x,t)\{ \frac{\partial \Drho(\x,t)}{\partial t}-D \nabla ^{2}\Drho( \x,t)\}
+D\rho_0 (\nabla i\hrho(\x,t) )^{2}\Big]-  \frac{1}{2}\int d^d\x \int d^d\x' \Phi(\x)\Delta^{-1}(\x-\x')\Phi(\x')
\end{equation}
and
\begin{equation}
\label{EffectiveAction_naive_v}
S^v_\Delta= \int d^d\x \int dt \Big[ D\Drho(\x,t) ( \nabla i\hrho(\x,t) )^{2}+\Gamma (i\hrho(\x,t)) \nabla\cdot\left \{ (\Drho(\x,t)+\rho_0)\nabla\Phi(\x) \right \} \Big] .
\end{equation}
\end{widetext}
The Gaussian part of the action $S^G_\Delta$ takes the similar form to $S_G$ in Eq.~(\ref{EffectiveActionG}). 
However, unlike $S^v$, $S^v_\Delta $ does not contain an infinite number of terms,
and takes a much simpler form. Here we regard
the last term in Eq.~(\ref{EffectiveAction_naive_v}), which is quadratic in fields as a part of vertices. 
As we will see below, due to the unusual nature of this vertex,
the perturbation expansion for this 
action is much more complicated than the corresponding scheme for $\delta\rho$
despite the apparent simplicity of the action.
We note that the functional integral over $ \Phi $ in Eq.~(\ref{EffectiveActionAvg_naive}) 
can actually be carried out to yield the effective action depending only on $ \Drho $ and $ \hrho $. This effective action
has been used in Ref.~\cite{kim2020dynamics}. For the present discussion, we find 
it more convenient to consider the $\Phi$-dependent action
as in Eq.~(\ref{EffectiveAction_naive_v}). We will pair up the dashed lines in the perturbation expansion as before,
which will produce the same diagrams as the method used in Ref.~\cite{kim2020dynamics}.

\begin{figure}
  \includegraphics[width=0.36\columnwidth]{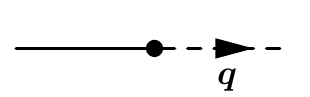} 
  \caption{Diagram for the type-$\mathcal{A}^0 $ vertex which exists only in the field theory for $\Delta\rho$.}
  \label{vertices_naive}
\end{figure}

We now develop the perturbation theory for the two-point function 
\begin{equation}
 F_{\alpha\beta}(\x-\x',t-t')= \langle \psi_\alpha(\x,t)\psi_\beta(\x',t')\rangle_{S_\Delta}
\end{equation}
evaluated with respect to $S_\Delta$ for the variables $\psi_\alpha, \psi_\beta$ representing
$\hat{\rho}$ or $\Delta\rho$ (denoted by $\rho$ 
in the subscript again). 
Since the Gaussian part $S^G_\Delta$ is identical to $S_G$ in Eq.~(\ref{EffectiveActionG}) except for $\Delta\rho$
and $\rho_0$ playing the roles of $\delta\rho$ and $\rho_0^\ast$, respectively, the bare propagators $F^0_{\alpha\beta}$
take the same form as the \textit{renormalized} $G^0_{\alpha\beta}$.
The noise vertex and the type-$ \mathcal{A} $ vertex in $S^v_\Delta$ have the same structure as before
with $\Delta\rho$ taking the place of $\delta\rho$.
Instead of the the type-$ \mathcal{B}_n $ vertex, the action
contains a new vertex as shown in Fig.~\ref{vertices_naive} which comes from the last term in 
Eq.~(\ref{EffectiveAction_naive_v}). We denote it by
the type-$\mathcal{A}_0 $ vertex. This carries the factor of
$- \rho_0 \Gamma \q^2$.

The type-$\mathcal{A}_0$ vertex contains just one $\hat{\rho}$ field. Therefore
the only way it can contribute to $F_{\rho\rho}$ is to appear at most twice in a diagram.
Otherwise, we would have too many $\hat{\rho}$'s to make a nonzero diagram. We can therefore classify the diagrams contributing to 
$ \tilde{F}_{\rho\rho} (\bm{q},t-t^\prime)$ into three distinct categories depending on the number of the new vertices in a diagram: 
(A) no type-$\mathcal{A}_0$ vertex, (B) one type-$\mathcal{A}_0$ vertex, and (C) two type-$\mathcal{A}_0$ vertices. 
The diagrams in category (A) are identical to the diagrams in (a) for $ G_{\rho\rho} $ discussed in the previous section. 

For diagrams in category (C), one cannot form a path from an external leg at time $t$ to the other one at time $t^\prime$ using only the solid (single or double) lines. The external legs have to be connected by a dashed line, which does not carry the time in it. This means that those in
category (C) give a constant contribution independent of $t-t^\prime$ (but still a function of $\bm{q}$). 
We shall explicitly show below that the contributions from the diagrams in category (B) reproduce those from the diagrams in (b) for $G_{\rho\rho}$ 
and that those in category (C) correspond to the time-independent term, Eq.~(\ref{DisconnectedCorrelation}), for $F_{\rho\rho}$. 

\begin{figure}
  \includegraphics{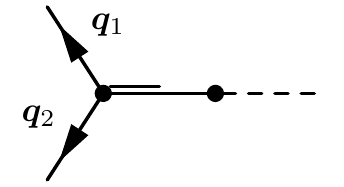} 
  \caption{Combination of the noise vertex and the type-$\mathcal{A}_0$ vertex which is equivalent to the renormalized
  type-$\mathcal{B}_1$ vertex.}
  \label{vertices_B1_form}
\end{figure}

\begin{figure}
  \includegraphics{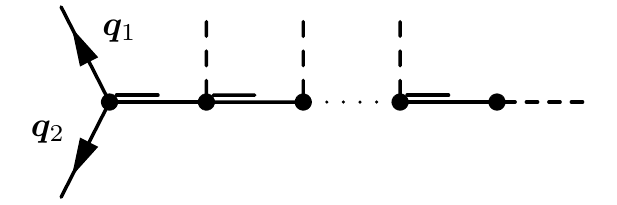} 
  \caption{Combination of the noise, type-$\mathcal{A}_0 $ and type-$\mathcal{A} $ vertices which is equivalent to the renornalized type-$ \mathcal{B}_n $ vertex.}
  \label{vertices_Bn_form}
\end{figure}

We investigate diagrams in category (B) in more detail. The diagrams in (B) must contain one noise vertex along
with one type-$\mathcal{A}^0 $ vertex. Figure~\ref{vertices_B1_form} shows the simplest 
vertex structure appearing in diagrams that belong to (B). 
We note that the external legs of this diagram have the same structure as the type-$\mathcal{B}_1$ vertex 
used in the previous section.
In the diagram in Fig.~\ref{vertices_B1_form}, the noise and the type-$\mathcal{A}_0$ vertices give the factors, $-D\bm{q}_1\cdot\bm{q}_2$ and
$-\rho_0\Gamma (\bm{q}_1+\bm{q}_2)^2$, respectively. Since one of the time integrals gives
\begin{equation}
\int_0^\infty dt\; i\tilde{F}^0_{\rho\hrho}(\bm{q}_1+\bm{q}_2,t) = \frac 1{D(\bm{q}_1+\bm{q}_2)^2},
\end{equation}
we end up with the overall factor of $\rho_0\Gamma \bm{q}_1\cdot \bm{q}_2$.
From Eq.~(\ref{typeB_renom}), we can see that this is exactly the factor for the renormalized type-$ \mathcal{B}_1 $ vertex.
Therefore, we can conclude that the diagram in Fig.~\ref{vertices_B1_form} plays exactly the same role as the 
renormalized type-$ \mathcal{B}_1 $ vertex.
In general, we can show that the diagram shown in Fig.~\ref{vertices_Bn_form} which consists of one noise vertex, 
one type-$ \mathcal{A}_0 $ vertex and $ n-1 $ type-$ \mathcal{A} $ vertices is equivalent to 
the renormalizd type-$ \mathcal{B}_n $ vertex shown in Fig.~\ref{vertices_renom}. 
The detailed proof for the equivalence is presented in Appendix \ref{app:b}. 
This can be regarded as one of the advantages in using the field theory for $\delta\rho$ over the corresponding one using $\Delta\rho$. 
A complicated combination of vertices in the field theory for $\Delta\rho$ can be represented as a single vertex in 
the corresponding formalism for $\delta\rho$. 

\begin{figure}
  \includegraphics{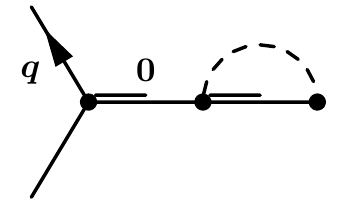} \;\; 
  \includegraphics{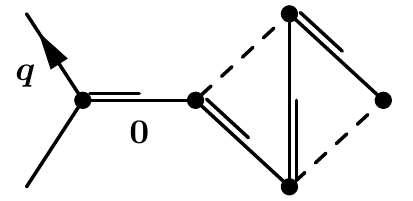} 
  \caption{Diagrams obtained by connecting the dashed lines in Fig.~\ref{vertices_Bn_form}. These kinds of diagrams vanish due to the incoming zero momentum. }
  \label{vertices_vanish}
\end{figure}

\begin{figure}
  \includegraphics{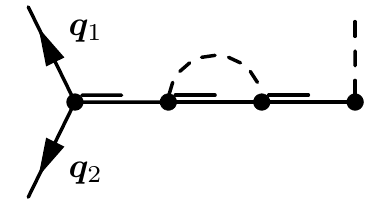} \;\; 
  \includegraphics{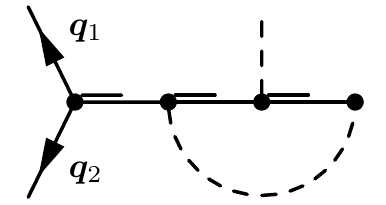} 
  \caption{Diagrams obtained by connecting the dashed lines in Fig.~\ref{vertices_Bn_form}.
  These diagrams cancel each other. }
  \label{vertices_B1_forms}
\end{figure}

We now consider the diagrams obtained by connecting the dashed lines in the diagram in Fig.~\ref{vertices_Bn_form}.
We recall that for the type-$\mathcal{B}_n$ vertex, connecting the dashed lines within a single vertex results in its renormalization.
But, as we have seen above, the diagram in Fig.~\ref{vertices_Bn_form} is already equivalent to the {\it renormalized} type-$\mathcal{B}_n$ vertex.
We therefore expect that the diagrams obtained by pairing up the dashed lines in Fig.~\ref{vertices_Bn_form} all vanish,
which we will demonstrate below. This means that 
the perturbation expansion for the $\Delta\rho$-field theory generates numerous unnecessary diagrams 
which as a whole give a vanishing contribution. 
We first note that when the two rightmost dashed lines are paired as in Fig.~\ref{vertices_vanish}, the momentum flowing through
the $\tilde{F}^0_{\rho\hat{\rho}}$ propagator right next to the loop must be zero. Then, since the type-$\mathcal{A}$ vertex involves the 
dot product between the zero momentum vector and another one, such a diagram vanishes.
When there is no such loop, a diagram does not vanish on its own in general. The diagrams in Fig.~\ref{vertices_B1_forms} show 
such examples. For these diagrams, when we perform the time integrals for the bare propagators, 
we obtain, apart from the common factor of $ (\rho_0 D /T^3)  \q_1 \cdot \q_2$, 
\begin{align}
\label{B1_formeq1}
   \int  \frac{d^d\q'}{(2\pi)^d}\;  \frac{(\q_1+\q_2)\cdot \q'}{(\q_1+\q_2)^2 } 
  \frac{ (\q_1+\q_2+\q')\cdot \q'}{(\q_1+\q_2+\q')^2} \tilde{\Delta}(q'), 
\end{align}
and
\begin{align}
\label{B1_formeq2}
  \int  \frac{d^d\q'}{(2\pi)^d}\;  \frac{(\q_1+\q_2)\cdot \q'}{(\q_1+\q_2)^2 } 
 \frac{ (\q_1+\q_2+\q')\cdot  (\q_1+\q_2) }{(\q_1+\q_2+\q')^2} \tilde{\Delta}(q'),
\end{align}
respectively. Here the momentum $\bm{q}^\prime$ flows through the dashed line loop.
We find that the individual diagrams do not vanish, but the sum does, due to the rotational invariance of the integration over $\bm{q}^\prime$.
We expect that for the higher order vertices equivalent to the type-$\mathcal{B}_n$, a similar cancellation must occur when the dashed lines are paired
within the vertex. This cancellation is a generic feature of the field theory involving $\Delta\rho$. We stress again that in the field theory for $\delta\rho$,
these kinds of unnecessary diagrams do not arise. These self-loops in these vertices are handled by the simple renormalization of 
the $\mathcal{B}_n$ vertex from the outset.

\begin{figure}
\includegraphics{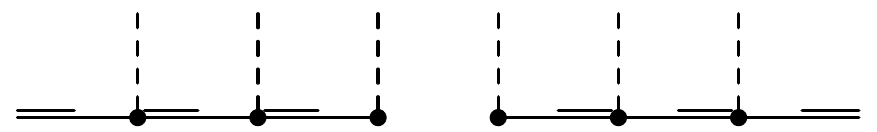} \\[8pt]
\caption{Typical diagram in category (C). Note that there must be a gap in the middle.
An example of six dashed lines is shown. For a general case, 
each disconnected part can contain an arbitrary number of dashed lines.
All possible pairs of the dashed lines are to be connected. }
\label{CGen}
\end{figure}

\begin{figure}
\includegraphics{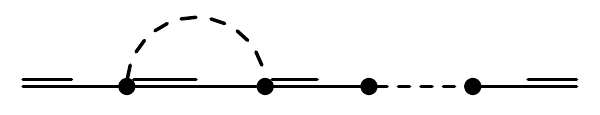} \\[8pt]
\includegraphics{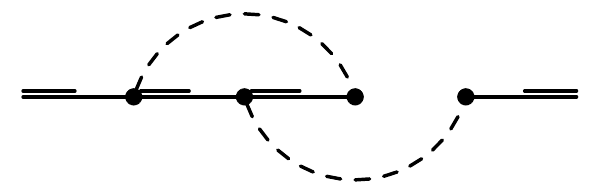} \\
\caption{Diagrams in category (C) that cancel each other. Compare this with Fig.~\ref{vertices_B1_forms}. }
\label{cat_C_cancel}
\end{figure}

As mentioned above, 
the diagrams in category (C) are responsible for the time-independent part given in Eq.~(\ref{DisconnectedCorrelation}).
This part arises only in this formalism, since $\Delta\rho$ is not a density fluctuation around its own average value,
but around the uniform value. We again find that many diagrams in the category (C) cancel each other and do not contribute at all.
We note that a generic diagram in (C) can be described schematically as in Fig.~\ref{CGen}. 
There is a gap in the middle between the two type-$\mathcal{A}_0$ vertices and
we can naturally split the diagram into two disconnected parts.
We then have to connect all possible pairs of the dashed lines. 
We find that a set of diagrams containing a dashed line connection within a disconnected part gives a vanishing contribution.
An example is shown in Fig.~\ref{cat_C_cancel}. We can easily see that these diagrams cancel each other, as
the diagrammatic structure is essentially the same as those in Fig.~\ref{vertices_Bn_form}.  
The nonvanishing diagrams are those with the dashed lines on the left part are connected 
to those on the right part. As we show in detail in Appendix \ref{app:c},
such a diagram with $2n$ dashed lines gives 
\begin{align}
\label{c_special} 
\frac{\rho_0^2}{n! T^{2n}} &\int \prod_{j=1}^{n}  \Big[ \frac{d^d\q_j}{(2\pi)^d} \tilde{\Delta}(\q_j) \Big]  (2\pi)^d \delta^{(d)} (\q-\sum_{i=1}^n\q_i ) ,
\end{align}
where $\bm{q}$ is the momentum flowing through the diagram in Fig.~\ref{CGen}. 
This is exactly the Fourier transformation of the right hand side of Eq.~(\ref{DisconnectedCorrelation}) at $O(\Delta^n)$.

In this section we have shown that, although the field theory for the density fluctuation around the uniform average 
is equivalent to that for $\delta\rho$
studied in the previous section,
its perturbation expansion generates numerous diagrams most of which vanish or cancel each other.
This could make the calculation of the correlation functions unnecessarily complicated if done in this formalism.

%%%%%%%%%%%%%%%%%%%%%%%%%%%%%%%%%%%%%%%%%

%\FloatBarrier
%\vspace{3mm}
\section{The physical response function and the fluctuation-dissipation relation}
\label{sec:resp}

In this section, we further explore the usefulness of the field theory for $\delta\rho$ 
developed in Sec.~\ref{sec:field} in the context of the FDR, which is obeyed by
the equilibrium dynamics described by the Langevin equations, Eqs.~(\ref{originalmodel}) and (\ref{DeanEquation}). The FDR,
which comes from the time-reversal symmetry of the equilibrium state,
provides the relationship between the correlation and the response functions. 
The MSRJD formalism is known to be suited for studying the
correlation and response of a system described by Langevin equations 
as the hatted variable arising in the field theory naturally provides the expression for the response function.
However, for the Langevin equation in Eq.~(\ref{DeanEquation}) given in terms of the density variable, the response function does not take
the simple form such as $\overline{\langle\rho(\x,t)i\hat{\rho}(\x',t')\rangle}$. 
It is well known 
that due to the multiplicative nature of the noise in Eq.~(\ref{DeanEquation}), 
the physical response to an external perturbation coupled to the density variable takes a more complicated form 
given by \cite{miyazaki2005mode,andreanov2006dynamical,kim2008fluctuation,kim2020dynamics} 
\begin{align}
\label{Phy_resp}
R(\x-\x',t-t')=- \Gamma \overline{ \langle \rho(\x,t)\nabla' \cdot (\rho(\x',t') \nabla' i \hrho(\x',t')) \rangle},
\end{align}
which we refer to as the physical response function in the following.
The FDR relates $R$ to the correlation function $C$ via
\begin{align}
\label{FDR_general}
- \frac{\partial}{\partial t} C(\x,t) = TR\left(\x,t \right) - TR \left (\x,-t \right),
\end{align}
where 
\begin{equation}
C(\bm{x}-\bm{x}',t-t')=\overline{\langle \rho(\bm{x},t)\rho(\bm{x}',t')\rangle}
\end{equation}
is the density-density correlation function. If we use $\rho(\bm{x},t)=\rho_\Phi(\bm{x})+\delta\rho(\bm{x},t)$, 
this is related to our correlation function as
\begin{equation}
C(\bm{x}-\bm{x}',t-t')=\overline{\rho_\Phi(\bm{x})\rho_\Phi(\bm{x}')}+G_{\rho\rho}(\bm{x}-\bm{x}',t-t').
\end{equation}

The physical response function in our field theory for $\delta\rho$ is obtained by simply 
replacing $ \rho(\x,t) $ in Eq.~(\ref{Phy_resp}) by $ \drho(\x,t)+\rho_{\Phi}(\x) $. The average is then given with respect to the effective
action $S$ in Eqs.~(\ref{EffectiveActionG}) and (\ref{EffectiveActionnG}). We first note that 
both $\langle \rho_\Phi(\x)\nabla' \cdot (\delta\rho(\x',t') \nabla' i \hrho(\x',t')) \rangle_S$ 
and $ \langle \drho(\x,t)\nabla' \cdot (\drho(\x',t') \nabla' i \hrho(\x',t'))\rangle_S $ are equal to zero in this field theory. 
The former is essentially $G_{\rho\hat{\rho}}$ evaluated at the same space time point. By connecting the end points in the generic diagram 
in Fig.~\ref{RespGen}, we find that we are left with the loops described in Fig.~\ref{loop}, which vanish due to causality.
The latter quantity must involve one noise vertex and can contain an arbitrary number of the type-$\mathcal{A}$ vertices. 
But, since there is no vertex 
with two $\delta\rho$ fields, the loops that appear in the perturbation expansion of this quantity are again all in the form
of Fig.~\ref{loop}.
Therefore, Eq.~(\ref{Phy_resp}) reduces to
\begin{align}
\label{Phy_resp_spec}
 R(\x-\x',t-t')
 =-\Gamma \left \langle \drho(\x,t)\nabla' \cdot (\rho_\Phi(\x') \nabla' i \hrho(\x',t')) \right \rangle_S. 
\end{align}
In the absence of disorder, $\rho_\Phi=\rho_0$, and the physical response function reduces to 
$\rho_0\Gamma \nabla'^2 iG^f_{\rho\hat{\rho}}(\x-\x',t-t')$ \cite{velenich2008brownian}, where the superscript $f$ denotes the case where 
$\Phi$ is set to zero. 
The effect of disorder in our field theory on the physical response function is 
reflected simply in the appearance of the inhomogeneous average density $\rho_\Phi$.

On the other hand, in the field theory using $\Delta\rho$, the physical response function takes a more involved form. By substituting $\rho$ with 
$\rho_0+\Delta\rho$ in Eq.~(\ref{Phy_resp}), we have
\begin{align}
\label{Phy_resp_naive}
 &R(\x-\x',t-t')=-\rho_0\Gamma \nabla'^2 iF_{\rho\hat{\rho}}(\x-\x',t-t') \nonumber \\
 &\quad\quad -\Gamma \left \langle \Delta\rho(\x,t)\nabla' \cdot (\Delta\rho(\x',t') \nabla' i \hrho(\x',t')) \right \rangle_{S_\Delta}. 
\end{align}
Evaluating the physical response function in this case involves the calculation of two correlation functions
with respect to the effective action $S_\Delta$, one of which is
a three-point function. The three-point function is nonvanishing in this case due to the presence of the type-$\mathcal{A}_0$ vertex.
In the absence of disorder, the three-point function in Eq.~(\ref{Phy_resp_naive}) vanishes, and we 
have only the first term and have the same result as before:
$\rho_0\Gamma \nabla'^2 iF^f_{\rho\hat{\rho}}(\x-\x',t-t')$ (note that $F^f_{\rho\hat{\rho}}=G^f_{\rho\hat{\rho}}$).
The effect of disorder in this case is encoded in the three-point function as well as in $F_{\rho\hat{\rho}}$
both of which have to be evaluated explicitly.

\begin{figure}
\includegraphics[width=0.72\columnwidth]{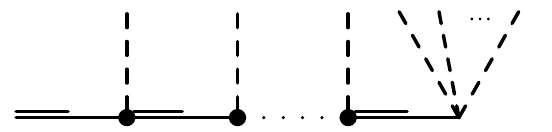} 
\caption{Generic diagram contributing to the physical response function $ R $. 
All possible pairs of the dashed lines are to be connected.}
\label{PhyRespGen}
\end{figure}

The simple structure of the physical response function, Eq.~(\ref{Phy_resp_spec}), in our field theory for $\delta\rho$ enables us to 
do a nonperturbative calculation. If we expand $\rho_\Phi$ in Eq.~(\ref{Phy_resp_spec}) in powers of $\Phi$,
we find that a generic diagram contributing to $R$ takes the form depicted in Fig.~\ref{PhyRespGen}, where
all possible pairs of dashed lines are to be connected. On the right end point $(\bm{x}',t')$, we have multiple $\Phi(\bm{x}')$'s coming from
the expansion of $\rho_\Phi$ along with $\hat{\rho}$ field. In the middle, we have dashed lines from a collection of 
the type-$\mathcal{A}$ vertices reaching up to $(\bm{x},t)$ where $\delta\rho$ lies.
The contribution from this kind of diagram with total $n$ dashed lines to the Fourier transform $\tilde{R}(\bm{q},t-t')$ 
can be written as
\begin{align}
\label{Xi_def}
\int \prod_{i=1}^n \Big[\frac{d^d \bm{k}_i}{(2\pi)^d} \; \tilde{\Phi}(\bm{k}_i)\Big]  \Xi_n (\bm{q};\bm{k}_1,\bm{k}_2,\cdots,\bm{k}_n;t-t')
\end{align}
for some vertex function $\Xi_n$ with the condition $\sum_{i=1}^n \bm{k}_i=0$. We note that, by construction, $\Xi_n$ is symmetric under the 
permutation of $n$ momenta, $(\bm{k}_1,\bm{k}_2,\cdots,\bm{k}_n)$. The response function $R$ can then be obtained by connecting all possible 
pairs of $\Phi$'s and by summing over all $n=0,1,2,\cdots$.

Despite the simple diagrammatic structure, 
it is still a difficult task to find a general nonperturbative expression for $\Xi_n$ and thus for $\tilde{R}(\bm{q},t-t')$. 
But we can make a progress if we focus on the zero frequency component of the response function,
\begin{equation}
\tilde{R}(\bm{q},\omega=0)=\int_{-\infty}^\infty dt\; \tilde{R}(\bm{q},t),
\end{equation}
where the time integral is actually from 0 to $\infty$ due to the causal nature of the response function. 
This quantity provides an important physical insight on the long-time behavior of the correlation function, since it is equal to
$ (\tilde{C}(\q,0)-\tilde{C}(\q,\infty))/T $ as can be obtained from the FDR, Eq.~(\ref{FDR_general}).
In order to calculate this, we define
\begin{align}
&\tilde{\Xi}_n (\bm{q};\bm{k}_1,\bm{k}_2,\cdots,\bm{k}_n;\omega=0) \nonumber \\
&~~~ \equiv\int_{-\infty}^\infty dt\; \Xi_n (\bm{q};\bm{k}_1,\bm{k}_2,\cdots,\bm{k}_n; t).
\label{Xi_def_int}
\end{align}
We evaluate this in detail in Appendix \ref{app:d}. The result is quite simple as it gives just a constant
\begin{align}
\label{XiRP_text}
\tilde{\Xi}_n(\q; \bm{k}_1, \cdots, \bm{k}_n;\omega=0 ) = -\frac{\rho^\ast_0 }{ (-T)^{n+1}}\frac{1}{n!}.
\end{align}
We now connect all possible pairs of $\Phi$'s in Eq.~(\ref{Xi_def}) and sum these contributions over $n=0,1,2,\cdots$.
It is clear that only the terms with even $n=2p$ survive. We therefore have
\begin{align}
\label{phy_resp_eq}
&\tilde{R}(\bm{q},\omega=0)=  \sum_{p=0}^{\infty}  \frac{(2p)!}{2^p p! }  \int \prod_{i=1}^{p} \frac{d^d\bm{k}_i}{(2\pi)^d} \tilde{\Delta}(\bm{k}_i) \\ 
 & \times \tilde{\Xi}_{2p} \left (\q; \bm{k}_1,-\bm{k}_1,\bm{k}_2,-\bm{k}_2, \cdots \bm{k}_p,-\bm{k}_p ;\omega=0 \right )   \nonumber
\end{align}
with the understanding that $p=0$ term is given just by $\Xi_0= \rho^\ast_0/T$. The factor $(2p)!/(2^p p!)$ accounts for the number of
possible ways to form all possible pairs of $\Phi$'s. Using Eq.~(\ref{XiRP_text}), we finally have
\begin{align}
\label{phy_resp_exact}
\tilde{R}(\q,\omega=0) &= \frac{\rho^\ast_0}{T}\sum_{p=0}^{\infty}  
\frac{1}{p!}  \left\{ \int \frac{d^d\bm{k}}{(2\pi)^d} \frac{\tilde{\Delta}(\bm{k}) }{2 T^2} \right \}^p  \nonumber \\
&= \frac{\rho^\ast_0}{T} \, e^{\Delta(\bm{0})/2T^2} = \frac{\rho_0}{T}.
\end{align}

%\FloatBarrier

As mentioned above, this nonperturbative result on the response function has an implication 
on the long-time behavior of the density-density correlation function. At time $t=0$, 
the particles are at equilibrium with respect to a given realization of the external potential $\Phi(\bm{x})$ at temperature $T$.
The density-density correlation function must be equal to the static one given by
\begin{align}
\langle\rho(\bm{x})\rho(\bm{x}')\rangle_{\text{st}}=\delta(\bm{x}-\bm{x}')\rho_\Phi(\bm{x})
+\rho_\Phi(\bm{x})\rho_\Phi(\bm{x}'), 
\end{align}
with $\rho_\Phi$ given in Eq.~(\ref{rhophi_2}).
Averaging over the disorder configurations, we therefore have
\begin{align}
C(\bm{x}-\bm{x}',0)=\delta(\bm{x}-\bm{x}')\rho_0
+\overline{\rho_\Phi(\bm{x})\rho_\Phi(\bm{x}')}.
\end{align}
The nonperturbative result, Eq.~(\ref{phy_resp_exact}), together with the FDR, Eq.~(\ref{FDR_general}) implies that
\begin{align}
C(\bm{x}-\bm{x}',\infty)=\overline{\rho_\Phi(\bm{x})\rho_\Phi(\bm{x}')}
\end{align}
or $G_{\rho\rho}(\bm{x}-\bm{x}',t)$ goes to zero in the long time limit. This means that the system remains ergodic and 
there is no ergodicity breaking transition for all values of
the disorder strength.

\FloatBarrier
%\vspace{3mm}
\section{Discussion and Conclusion}

We have constructed the MSRJD dynamical field theory for the noninteracting Brownian particles 
in a quenched Gaussian random potential. We have set up the diagrammatic perturbation scheme for
the connected density-density correlation function. The main variable for our field theory is the density fluctuation
around the nonuniform average value. The diagrammatic structures are compared in detail with those in
the field theory using the density fluctuation around the uniform average value, which is 
used in Ref.~\cite{kim2020dynamics}. It is shown that the latter generates many unnecessary diagrams
which either vanish or cancel among themselves. Using our field theory, we were able to 
evaluate the zero-frequency component of the response function exactly by summing all the diagrams.
According to the FDR, our result implies that the connected density-density correlation function 
decays to zero in the long time limit and the system remains ergodic for all values of the disorder strength.

Our finding is in contrast to that was found in Ref.~\cite{kim2020dynamics}. In that paper, various renormalized 
perturbation schemes for the density correlation function, which are consistent with the FDR, are presented. In one of the 
schemes, the connected density correlation function does not decay to zero in the long-time limit,
but approaches a finite value when the disorder strength exceeds some critical value. This signals an ergodic-nonergodic transition.
This is, however, in contrast to our exact result, since if the transition exists, the right hand side
of Eq.~(\ref{phy_resp_exact}) would give a different value at the transition. These renormalized perturbation theories
basically correspond to replacing the bare correlation functions in some perturbation expansion scheme 
with the renormalized ones. These in turn produce various types of self-consistent equations for the renormalized correlation function.
In terms of the Feynman diagrams, a renormalized perturbation theory corresponds to the partial resummation 
of a particular infinite subset of diagrams contributing to the density correlation function.
Our exact result suggests that the ergodic-nonergodic transition seen in Ref.~\cite{kim2020dynamics}
might be an artifact of the partial resummation of the diagrams.  The situation is very reminiscent of 
the dynamical transition predicted by the MCT of supercooled liquids \cite{gotze2008complex}, which can also be regarded
as a partial resummation of diagrams for the full theory of supercooled liquids.  
It is expected that the sharp transition will be smeared out when other effects such as activated hopping are included 
and that the system remains ergodic \cite{das1986fluctuating,schmitz1993absence,das2004mode}. 

In the present work, we have only considered the zero-frequency component of the response function. In order to find more useful
information on the transport properties of the system it will be necessary to find 
a reliable scheme to calculate the full time dependence of the correlation and response functions.
We have tried to implement various renormalized perturbation schemes
including those presented in Ref.~\cite{kim2020dynamics} for the density correlation function.
We have also encountered the same problems as in Ref.~\cite{kim2020dynamics}
such as spurious instabilities when we try to find solutions to self-consistent equations.
We believe that in order to further improve the renormalized perturbation 
theory for this system, one needs to consider the renormalization of the vertices as well as the propagators and
to find self-consistent equations for these quantities. This is left for future work. 
For this kind of calculations and also for an eventual generalization to the system of interacting Brownian particles in a quenched
random potential, we believe that the perturbation scheme presented in this work 
would provide a convenient starting point.

\FloatBarrier
%\vspace{3mm}
\begin{widetext}

\appendix
\section{Properties of Gaussian disorder average}
\label{app:a}

We consider the disorder average of the denominator on the right hand side of Eq.~(\ref{rhophi}). We have
\begin{align}
\overline{\int d^d\bm{x}\;e^{-\Phi(\bm{x})/T} }=\int d^d\bm{x}\; e^{\Delta(\bm{0})/(2T^2)}=Ve^{\Delta(\bm{0})/(2T^2)},
\end{align}
where $V$ is the volume.
We calculate the variance of this quantity as
\begin{align}
\overline{ \left[ \int d^d\bm{x}\;e^{-\Phi(\bm{x})/T} - Ve^{\Delta(\bm{0})/(2T^2)} \right]^2} 
=& \int d^d\bm{x} \int d^d\bm{x}'\; e^{(\Delta(\bm{0})+\Delta(\bm{x}-\bm{x}'))/T^2}-V^2e^{\Delta(\bm{0})/T^2} \nonumber \\
=&Ve^{\Delta(\bm{0})/T^2} \int d^d\bm{x} \;(e^{\Delta(\bm{x})/T^2}-1).
\end{align}
If we assume $\Delta(\bm{x})$ is a short-ranged function, then the integral gives a finite quantity.
Therefore the disorder average and the variance of $\int d^d\bm{x}\;e^{-\Phi(\bm{x})/T}$ are both proportional to $V$
and we can replace it by its average value in the thermodynamic limit.

\section{Evaluation of the diagram in Fig.~\ref{vertices_Bn_form}}
\label{app:b}

We consider the case where there are $n$ dahsed lines in Fig.~\ref{vertices_Bn_form} carrying 
the momenta $ \bm{k}_1, \bm{k}_2, \cdots \bm{k}_n $. Figure \ref{vertices_Bn_form} can be represented in the action as
\begin{align}
\label{vertex_int}
\begin{split}
\int dt & \int \frac{d^d\bm{q}_1}{(2\pi)^d}\int \frac{d^d\bm{q}_2}{(2\pi)^d}\int\prod_{i=1}^n \frac{d^d\bm{k}_i}{(2\pi)^d}\;
(2\pi)^d\delta^{(d)}(\bm{q}_1+\bm{q}_2-\sum_{i=1}^n \bm{k}_i)  \\ 
&\times (-D  \q_1 \cdot \q_2) \Lambda_n (\bm{k}_1,\bm{k}_2,\cdots,\bm{k}_n) i\hat{\rho}(-\bm{q}_1,t)
i\hat{\rho}(-\bm{q}_2,t) 
\prod_{i=1}^n \tilde{\Phi}(\bm{k}_i) 
\end{split}
\end{align}
for some vertex function $\Lambda_n$. In order to evaluate $\Lambda_n$, we consider the case where
the momenta going in through the dashed lines from right to left in Fig.~\ref{vertices_Bn_form} are given by $ \bm{k}_1, \bm{k}_2, \cdots \bm{k}_n $.
For convenience, we use the shorthand notation for the bare propagator as
\begin{equation}
g(\bm{q},t)\equiv i\tilde{G}^0_{\rho\hat{\rho}}(\bm{q},t)= i\tilde{F}^0_{\rho\hat{\rho}}(\bm{q},t)=\Theta(t)e^{-D\bm{q}^2t},
\label{bare_short}
\end{equation}
Then, for this particular configuration of the momenta, the contribution to $\Lambda_n$ is given by 
\begin{align}
\begin{split}
  \int_{-\infty}^\infty & \prod_{i=1}^{n}dt_i \; 
\{ -\rho_0\Gamma \bm{k}_1\cdot\bm{k}_1\} g(\bm{k}_1,t_2-t_1) 
\{-\Gamma \bm{k_2}\cdot (\bm{k}_1+\bm{k}_2)\} g(\bm{k}_1+\bm{k}_2,t_3-t_2)\\
\times \cdots &\times \{-\Gamma \bm{k}_{n-1}\cdot(\bm{k}_1+\cdots+\bm{k}_{n-1})\}
g(\bm{k}_1+\cdots+\bm{k}_{n-1},t_{n}-t_{n-1}) \\
&\times \{-\Gamma \bm{k}_n\cdot(\bm{k}_1+\cdots+\bm{k}_n)\}
g(\bm{k}_1+\cdots+\bm{k}_n,t-t_n),
\end{split}
\end{align}
where $t_i$ denotes the time when the momentum $\bm{k}_i$ is coming in for $i=1,2,\cdots,n$. 
We make the change of variables,
$s_i=t_{i+1}-t_i$ for $i=1,\cdots,n$, ($t_{n+1}=t$) with the unit Jacobian. Then the time integral over $s_i$ is from 0 to $\infty$ due to
the step function in the bare propagator and can be evaluated explicitly as
\begin{align}
\label{ver_n_form}
\rho_0    \prod_{i=1}^{n}   \left\{  -  \Gamma \, \bm{k}_i \cdot \left( \bm{k}_1+\bm{k}_2 +\cdots \bm{k}_i \right) 
\int_0^\infty ds_i \; e^{-D \left( \bm{k}_1+\bm{k}_2 +\cdots \bm{k}_i \right)^2 s_i }  \right\}
= \frac{ \rho_0}{(-T)^n}  
\prod_{i=1}^{n}    \frac{ \bm{k}_i \cdot \left( \bm{k}_1+\bm{k}_2 +\cdots \bm{k}_i \right)}{ \left( \bm{k}_1+\bm{k}_2 +\cdots \bm{k}_i \right)^2}  ,
\end{align}
As can be seen from Eq.~(\ref{vertex_int}), $\Lambda_n$ can be symmetrized with respect to the permutation of its arguments. Therefore, we have
\begin{equation}
\Lambda_n(\bm{k}_1,\bm{k}_2,\cdots,\bm{k}_n)=\frac 1{n!}\frac{ \rho_0}{(-T)^n} O_n (\bm{k}_1,\bm{k}_2,\cdots,\bm{k}_n),
\label{LO}
\end{equation}
where
\begin{align}
\label{O_function}
O_n(\bm{k}_1, \bm{k}_2, \cdots \bm{k}_n) \equiv \sum_{P}    \prod_{i=1}^{n}   
 \frac{ \bm{k}_i \cdot \left( \bm{k}_1+\bm{k}_2 +\cdots \bm{k}_i \right)}{ \left( \bm{k}_1+\bm{k}_2 +\cdots \bm{k}_i \right)^2}
\end{align}
with the summation over the permutations $P$ of $(\bm{k}_1, \bm{k}_2, \cdots \bm{k}_n)$.

We now prove by induction that $O_n=1$ for all $n=1,2,\cdots$. It is trivial to see that $O_1=1$ and that
\begin{align}
\label{O_function_2}
 O_2(\bm{k}_1,\bm{k}_2)=  \frac{\bm{k}_2 \cdot  (\bm{k}_1+\bm{k}_2)  }{ (\bm{k}_1+\bm{k}_2)^2  } \frac{ \bm{k}_1^2}{ \bm{k}_1^2} + \frac{\bm{k}_1 \cdot  (\bm{k}_1+\bm{k}_2)  }{ (\bm{k}_1+\bm{k}_2)^2  } \frac{ \bm{k}_2^2}{ \bm{k}_2^2} = 1.
\end{align}
Now we suppose that  $ O_{n-1}=1 $. We note that, compared to $O_{n-1}$, $O_n$ contains one additional factor which contains all the $\bm{k}_i$'s. 
Therefore, we can write 
\begin{align}
\label{O_function_recursive}
O_{n}(\bm{k}_1, \bm{k}_2, \cdots, \bm{k}_{n})  = \sum_{i=1}^{n}  
 \frac{\bm{k}_i \cdot \left( \bm{k}_1+\bm{k}_2 +\cdots \bm{k}_{n} \right)}{ \left( \bm{k}_1+\bm{k}_2 +\cdots \bm{k}_{n} \right)^2} 
 O_{n-1}(\bm{k}_1, \cdots, \bm{k}_{i-1},\bm{k}_{i+1},\cdots,\bm{k}_n)  ,
\end{align}
and from the assumption we conclude that
\begin{align}
\label{O_function_solve}
O_{n}(\bm{k}_1, \bm{k}_2, \cdots, \bm{k}_{n})  =\sum_{i=1}^{n}    \frac{\bm{k}_i \cdot \left( \bm{k}_1+\bm{k}_2 +\cdots \bm{k}_{n} \right)}{ \left( \bm{k}_1+\bm{k}_2 +\cdots \bm{k}_{n} \right)^2} =1.
\end{align}
Therefore, from Eq.~(\ref{LO}), we have $\Lambda_n=\rho_0/(n! (-T)^n)$. By using this value in Eq.~(\ref{vertex_int}) and comparing
it with the last term in Eq.~(\ref{EffectiveActionnG}), we find that
the diagram in Fig.~\ref{vertices_Bn_form} is equal to the \textit{renormalized} type-$\mathcal{B}_n$ vertex, which carries $\rho_0$ instead of $\rho^\ast_0$.

\section{Evaluation of the diagram in Fig.~\ref{CGen}}
\label{app:c}

We first note that both disconnected parts in Fig.~\ref{CGen} have the same structure as the diagram in Fig.~\ref{vertices_Bn_form}
treated in Appendix~\ref{app:b} except that at the end points we have the $\Delta\rho$ fields instead of the noise vertex.
For the disconnected part on the left hand side, if we denote the external momentum coming out of the left end point by $\bm{q}$ and 
the momenta coming in the $n$-dashed lines by $\bm{k}_i$, $i=1,2,\cdots,n$, we can write this part as
\begin{align}
\int\prod_{i=1}^n \frac{d^d\bm{k}_i}{(2\pi)^d}\; (2\pi)^d\delta^{(d)}(\bm{q}-\sum_{i=1}^n \bm{k}_i) 
 \Lambda_n (\bm{k}_1,\bm{k}_2,\cdots,\bm{k}_n) \prod_{i=1}^n \tilde{\Phi}(\bm{k}_i) 
\end{align}
with the same vertex function $\Lambda_n$ as considered in Eq.~(\ref{vertex_int}). As expected, there is no time dependence. 
We represent
the part on the right hand side in the same way using the momenta $\bm{q}^\prime$ and $\bm{k}^\prime_i$, $i=1,2,\cdots,n$.
We then connect all possible pairs of the dashed lines. This produces the disorder correlation $\tilde{\Delta}(\bm{k}_i)$ 
with the delta function enforcing $\bm{k}_i+\bm{k}^\prime_i=0$.
Therefore, along with the overall delta function $(2\pi)^d\delta^{(d)}(\bm{q}+\bm{q}^\prime)$, we have
\begin{align}
\label{c_special_0}
n!\int \prod_{i=1}^{n}  \left ( \frac{d^d\bm{k}_i}{(2\pi)^d} \tilde{\Delta}(\bm{k}_i) \right)  \; \Lambda_n^2 (\bm{k}_1,\bm{k}_2,\cdots \bm{k}_{n})
 (2\pi)^d \delta^{(d)} (\q-\sum_{i=1}^n\bm{k}_i),
\end{align}
where $n!$ denotes the number of different ways of connecting the dashed lines.
If we use the result $\Lambda_n=\rho_0/(n! (-T)^n)$ obtained in Appendix \ref{app:b},
we find that this is exactly Eq.~(\ref{c_special}).

\section{Evaluation of the vertex function $\tilde{\Xi}_n (\q; \bm{k}_1,\bm{k}_2,\cdots,\bm{k}_n ;\omega=0) $}
\label{app:d}

\begin{figure}
\includegraphics[width=0.4\columnwidth]{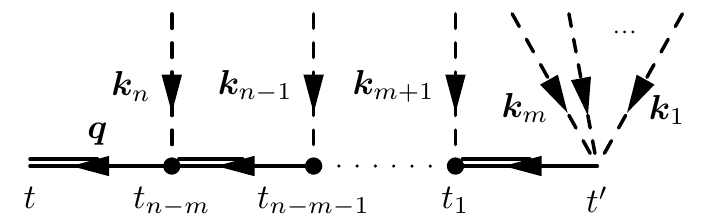} 
\caption{Diagram in  Fig.~\ref{PhyRespGen} with the mometum labels. This is evaluated in Appendix \ref{app:d}}
\label{PhyRespGenLabel}
\end{figure}

We consider the diagram in Fig.~\ref{PhyRespGen} with total $n$ dashed lines. 
Among the $n$ dashed lines, we consider the situation described in Fig.~\ref{PhyRespGenLabel}, where
there are $ m $ ($m\le n$) dashed lines at the right external source at time $t'$ and
$ n-m $ dashed lines coming from a series of the
type-$\mathcal{A}$ vertices reaching at the left external source at time $t$. 
We denote the $ m $ momenta coming in through the dashed lines at the right end point by $ \bm{k}_1, \bm{k}_2, \cdots \bm{k}_m $,
and those coming in at times 
$t_1, t_2, \cdots, t_{n-m}$ through the remaining dashed lines by $ \bm{k}_{m+1}, \bm{k}_{m+2}, \cdots \bm{k}_n $  from right to left, respectively. 
For $\tilde{R}(\bm{q},t-t')$, the momentum coming out of 
the left external source at time $t$ is $\bm{q}$ and the momentum conservation gives the condition that
that $\sum_{i=1}^n \bm{k}_i =0$.  For this particular configuration of $\{\bm{k}_i\}$, the contribution from this diagram to 
$\Xi_n (\bm{q}; \bm{k}_1,\bm{k}_2,\cdots,\bm{k_n};t-t')$ in Eq.~(\ref{Xi_def}) is 
\begin{align}
\begin{split}
&-\frac{\rho^\ast_0 \Gamma}{m! (-T)^m}\int_{-\infty}^\infty \prod_{i=1}^{n-m} dt_i\; \Big\{-\bm{q}\cdot(\bm{q}+\sum_{j=1}^m\bm{k}_j)\Big\} 
g\Big(\bm{q}+\sum_{j=1}^m\bm{k}_j,t_1-t'\Big) \\
&\times \Big\{-\Gamma\bm{k}_{m+1}\cdot(\bm{q}+\sum_{j=1}^{m+1}\bm{k}_j)\Big\}
g\Big(\bm{q}+\sum_{j=1}^{m+1}\bm{k}_j,t_2-t_1\Big)
\Big\{-\Gamma\bm{k}_{m+2}\cdot(\bm{q}+\sum_{j=1}^{m+2}\bm{k}_j)\Big\}
g\Big(\bm{q}+\sum_{j=1}^{m+2}\bm{k}_j,t_3-t_2\Big) \\
&\times\cdots\times \Big\{-\Gamma\bm{k}_{n}\cdot(\bm{q}+\sum_{j=1}^{n}\bm{k}_j)\Big\}
g\Big(\bm{q}+\sum_{j=1}^{n}\bm{k}_j,t-t_{n-m}\Big)
\end{split}
\end{align} 
with the bare propagator $g$ given in Eq.~(\ref{bare_short}). In general it is difficult to evaluate the time integrals to get a
closed form. But, if we focus on the time integral over $\tau=t-t'$ of the above quantity to get the zero-frequency 
limit of the response function from Eq.~(\ref{Xi_def_int}), we can make a progress.
By changing the integration variables from $t_1, \cdots, t_{n-m}$ and $\tau$ to $s_1=t_1-t'$, $s_2=t_2-t_1, \cdots , s_{n-m}=t_{n-m}-t_{n-m-1}$
and $s_{n-m+1}=t-t_{n-m}$, with the unit Jacobian, we can explicitly evaluate all the time integrals as we have done in Appendix \ref{app:b}.
The result is 
\begin{align}
\label{phyresp_form1}
 -\frac{\rho^\ast_0 }{m! (-T)^{n+1}} \frac{ \q \cdot \left(\q+ \bm{k}_1 +\cdots +\bm{k}_m  \right)}
 { \left( \q+ \bm{k}_1 +\cdots +\bm{k}_m  \right)^2} \prod_{i=m+1}^{n}     
 \frac{ \bm{k}_i \cdot \left(\q+ \bm{k}_1+\cdots+ \bm{k}_i  \right)}{ \left(\q+ \bm{k}_1+\cdots \bm{k}_i \right)^2}  .
\end{align}

As can be seen from Eq.~(\ref{Xi_def}), we have to symmetrize this quantity over all the permutations of $(\bm{k}_1,\cdots,\bm{k}_n)$.
We also have to consider all possible cases of $m=0,1,\cdots,n$ for given $n$. Therefore we can express the quantity 
in Eq.~(\ref{Xi_def_int}) as
\begin{align}
\label{Xi_result}
\tilde{\Xi}_n(\bm{q};\bm{k}_1,\bm{k}_2,\cdots,\bm{k}_n;\omega=0)=-\frac{\rho^\ast_0 }{(-T)^{n+1}} \frac 1 {n!} \sum_{m=0}^n 
O_{nm} (\bm{q};\bm{k}_1,\bm{k}_2,\cdots,\bm{k}_n),
\end{align}
where 
\begin{align}
\label{P_function}
O_{nm}(\q;\bm{k}_1, \bm{k}_2, \cdots, \bm{k}_n) \equiv \sum_{P}  \Biggl \{   \frac{1}{m!} \frac{ \q \cdot \left( \q+\bm{k}_1 +\cdots+ \bm{k}_m  
\right)}{\left( \q+ \bm{k}_1 +\cdots + \bm{k}_m  \right)^2} \prod_{i=m+1}^{n}   
 \frac{ \bm{k}_i \cdot \left( \q+\bm{k}_1+\cdots +\bm{k}_i \right)}{ \left(\q+ \bm{k}_1+\cdots +\bm{k}_i \right)^2}   \Biggr \},
\end{align}
where $\sum_P$ indicates the sum over all
possible permutations of $(\bm{k}_1, \bm{k}_2, \cdots, \bm{k}_n) $. 
We will prove below that $\sum_{m=0}^n O_{nm}=1$ for all $n=0,1,2,\cdots$ by mathematical induction.  First we note that $O_{00}=1$ and
\begin{align}
O_{10}(\bm{q};\bm{k}_1)=\frac{\bm{k}_1\cdot (\bm{q}+\bm{k}_1)}{(\bm{q}+\bm{k}_1)^2},~~~
O_{11}(\bm{q};\bm{k}_1)=\frac{\bm{q}\cdot (\bm{q}+\bm{k}_1)}{(\bm{q}+\bm{k}_1)^2}, 
\end{align}
thus $O_{10}+O_{11}=1$. Now we suppose that $\sum_{m=0}^{n-1}O_{n-1,m}=1$.  
Using the method similar to the one used to derive Eq.~(\ref{O_function_recursive}), we obtain
\begin{align}
\label{P_function_recursive}
O_{nm}(\q; \bm{k}_1, \bm{k}_2, \cdots, \bm{k}_{n})  = 
\sum_{i=1}^{n} \Big\{   \frac{\bm{k}_i \cdot \left( \q+ \bm{k}_1 +\cdots +\bm{k}_{n}\right)}
{ \left( \q+ \bm{k}_1+\cdots+ \bm{k}_{n}  \right)^2} O_{n-1, m}(\q; \bm{k}_1, \cdots, \bm{k}_{i-1},\bm{k}_{i+1},\cdots,\bm{k}_{n})  \Big\},
\end{align}
for $ m \leq n-1 $. We therefore find that
\begin{align}
\label{P_function_calculation}
\begin{split}
\sum_{m=0}^{n}  O_{nm}(\q; \bm{k}_1, \cdots, \bm{k}_{n}) = &\sum_{m=0}^{n-1}  O_{nm}(\q; \bm{k}_1, \cdots,  \bm{k}_{n}) 
+ O_{nn}(\q; \bm{k}_1, \cdots, \bm{k}_{n})   \\
=  & \sum_{m=0}^{n-1} \sum_{i=1}^{n} \Big\{   \frac{\bm{k}_i \cdot \left( \q+\bm{k}_1 +\cdots+ \bm{k}_{n}\right)}
{ \left(\q+ \bm{k}_1+\cdots \bm{k}_{n} \right)^2}  
O_{n-1, m}(\q; \bm{k}_1, \cdots, \bm{k}_{i-1},\bm{k}_{i+1},\cdots,\bm{k}_n)  \Big\} \\
&+ O_{nn}(\q,\{\bm{k}_1, \cdots \bm{k}_{n}\})  \\
=& \sum_{i=1}^{n}   \frac{\bm{k}_i \cdot \left( \q+\bm{k}_1 +\cdots+ \bm{k}_{n} \right)}{ \left( \q+\bm{k}_1+\cdots+ \bm{k}_{n}  \right)^2} 
+ \frac{ \q \cdot \left( \q+\bm{k}_1 +\cdots +\bm{k}_n  \right)}{\left( \q+\bm{k}_1 +\cdots+ \bm{k}_n \right)^2} = 1.  
\end{split}
\end{align}
From Eq.~(\ref{Xi_result}), we finally have
\begin{align}
\label{XiRP}
\tilde{\Xi}_n(\q; \bm{k}_1, \cdots, \bm{k}_n;\omega=0 ) = -\frac{\rho^\ast_0 }{ (-T)^{n+1}}\frac{1}{n!}
\end{align}
regardless of its arguments.

\end{widetext}

\begin{acknowledgments}
We would like to thank Bongsoo Kim for helpful discussion.
This work was supported by the National Research Foundation of Korea(NRF) 
grant funded by the Korea government(MSIT) (No.\ 2020R1F1A1062833), and 
by Konkuk University's research support program for its faculty on sabbatical leave in 2020.
\end{acknowledgments}

\bibliography{brownian_dis}

\end{document}